\AtBeginDocument{%
  }
    
\documentclass[sigconf]{acmart}
\usepackage{algorithmic}
\usepackage{graphicx}
\usepackage{textcomp}
\usepackage{xcolor}
\usepackage{hyperref}
\usepackage{soul}
\usepackage{physics,array}
\usepackage{multirow}
\usepackage{tikz}
\newcommand*\circled[1]{\tikz[baseline=(char.base)]{
            \node[shape=circle,fill,inner sep=0.5pt] (char) {\textcolor{white}{#1}};}}

\setcopyright{acmlicensed}
\copyrightyear{2018}
\acmYear{2018}
\acmDOI{XXXXXXX.XXXXXXX}

\renewcommand\footnotetextcopyrightpermission[1]{}

\acmBooktitle{Woodstock '18: ACM Symposium on Neural Gaze Detection,
 June 03--05, 2018, Woodstock, NY}
\acmISBN{978-1-4503-XXXX-X/18/06}

\begin{document}
\fancyhead{}

\title{MITS: A Quantum Sorcerer’s Stone For Designing Surface Codes}


\author{Avimita Chatterjee}
\email{amc8313@psu.edu}
\orcid{1234-5678-9012}
\affiliation{%
  \institution{Pennsylvania State University}
  \city{State College}
  \state{PA}
  \country{USA}
}

\author{Debarshi Kundu}
\affiliation{%
  \institution{Pennsylvania State University}
  \city{State College}
  \state{PA}
  \country{USA}
  }
\email{dqk5620@psu.edu}

\author{Swaroop Ghosh}
\affiliation{%
  \institution{Pennsylvania State University}
  \city{State College}
  \state{PA}
  \country{USA}
}
\email{szg212@psu.edu}


\begin{abstract}
      In the evolving landscape of quantum computing, determining the most efficient parameters for Quantum Error Correction (QEC) is paramount. Various quantum computers possess varied types and amounts of physical noise. Traditionally, simulators operate in a forward paradigm, taking parameters such as distance, rounds, and physical error to output a logical error rate. However, usage of maximum distance and rounds of the surface code might waste resources. An approach that relies on trial and error to fine-tune QEC code parameters using simulation tools like STIM can be exceedingly time-consuming. Additionally, daily fluctuations in quantum error rates can alter the ideal QEC settings needed. As a result, there is a crucial need for an automated solution that can rapidly determine the appropriate QEC parameters tailored to the current conditions. To bridge this gap, we present MITS, a tool designed to reverse-engineer the well-known simulator STIM for designing QEC codes. MITS accepts the specific noise model of a quantum computer and a target logical error rate as input and outputs the optimal surface code rounds and code distances. This guarantees minimal qubit and gate usage, harmonizing the desired logical error rate with the existing hardware limitations on qubit numbers and gate fidelity. We explored and compared multiple heuristics and machine learning models for training/designing MITS and concluded that XGBoost and Random Forest regression were most effective, with Pearson correlation coefficients of $0.98$ and $0.96$ respectively.
\end{abstract}





\maketitle

\section{Introduction} \label{sec:introduction}

Quantum Error Correction (QEC) is an indispensable pillar in the advancement of quantum computing. In the quantum realm, where information resides in fragile quantum states, even trivial interactions with the external environment can introduce errors, thereby threatening computational integrity \cite{mouloudakis2021entanglement}. Quantum systems are uniquely susceptible to two main types of errors: bit-flip and phase-flip. While classical systems often contend with bit-flip errors, the quantum realm introduces the added challenge of phase-flip errors. Classical error correction schemes are ineffective for quantum information due to multiple reasons \cite{wootters2009no, von2018mathematical}, necessitating QEC. Surface codes \cite{dennis2002topological} address both bit and phase errors by using a two-dimensional qubit layout. Such an approach makes them a frontrunner for near-term fault-tolerant quantum computing.  Within this category, there are `rotated' \cite{tomita2014low} and `unrotated' \cite{fowler2012surface} variants. Rotated surface codes offer a higher resistance to errors and are somewhat less complex to implement than the unrotated versions \cite{chatterjee2023q}. The need for QEC becomes increasingly crucial as we aim for larger, more reliable quantum systems. It is the key to ensure that the quantum computations are both accurate and consistent.

\textbf{Motivation:}
Quantum computing advancements prompt a critical inquiry: Does the expansion of error-correcting codes consistently reduce logical errors in practical applications? Consider, for example, a distance $25$ rotated surface code requiring roughly $580$ qubits and $2040$ gates, which approaches fault tolerance. However, the necessity for such a large-scale setup is questionable. Theoretically, larger codes are expected to lower logical errors, but QECC entails significant trade-offs, such as the need for more qubits and longer computation times, potentially heightening decoherence risks \cite{kelly2015state}. The effectiveness of QECCs also depends on the specific environmental noise they address. Thus, selecting the optimal QECC is not merely about reducing errors but also about customizing the code to match the physical error characteristics of the quantum computer in question.

\begin{table}[]
\centering
\caption{Surface Code Parameters and Overhead for Lowering Logical Error in Three Random `ibm\_osaka' Instances.}
\begin{tabular}{cc||ccc}
\multicolumn{2}{c||}{\textbf{Category}}                                                                                               & \textbf{Inst. 1} & \textbf{Inst. 2} & \textbf{Inst. 3} \\ \hline \hline
\multicolumn{1}{c|}{\multirow{3}{*}{\textbf{\begin{tabular}[c]{@{}c@{}}Physical\\ Error\\ Rates\end{tabular}}}}  & \textbf{Depol}    & 2.4E -4          & 2.1E -4          & 2.3E -4          \\ \cline{2-2}
\multicolumn{1}{c|}{}                                                                                            & \textbf{Gate}     & 7.7E -3          & 6.7E -3          & 8.4E -3          \\ \cline{2-2}
\multicolumn{1}{c|}{}                                                                                            & \textbf{Readout}  & 2.5E -2          & 2.1E -2          & 2.4E -2          \\ \hline
\multicolumn{2}{c||}{\textbf{Target Logical Error Rate}}                                                                              & 1E -9            & 1E -9            & 1E -9            \\ \hline \hline
\multicolumn{1}{c|}{\multirow{2}{*}{\textbf{\begin{tabular}[c]{@{}c@{}}Surface Code\\ Parameters\end{tabular}}}} & \textbf{Distance} & 21               & 23               & 25               \\ \cline{2-2}
\multicolumn{1}{c|}{}                                                                                            & \textbf{Rounds}   & 67               & 69               & 70               \\ \hline
\multicolumn{1}{c|}{\multirow{2}{*}{\textbf{Overhead}}}                                                          & \textbf{Qubits}   & $\sim 477$              & $\sim 528$              & $\sim 580$              \\ \cline{2-2}
\multicolumn{1}{c|}{}                                                                                            & \textbf{Gates}    & $\sim 1592$             & $\sim 1816$             & $\sim 20240$            \\ \hline
\multicolumn{2}{c||}{\textbf{Usual Simulation Time}}                                                                                  & $\sim 45$ hr          & $\sim 53$ hr          & $\sim 62$ hr          \\ \hline \hline
\end{tabular}
\label{tab:ibm_osaka_table}
\vspace{-20pt}
\end{table}

Furthermore, quantum computers undergo routine calibration, leading to fluctuations in physical error rates. This may require frequent adjustments to surface code parameters to meet the target error rate or save expensive resources. Over six weeks, we collected error readings from `ibm\_osaka' quantum computer. From this dataset, showcased in Table \ref{tab:ibm_osaka_table}, are three random instances to identify the precise surface code parameters needed to decrease the logical error rate to $10^{-9}$. Although the physical error values showed minimal variation, significant differences were observed in the required overhead for surface code parameters. For example, while `instance 1' necessitated a code distance of $21$, `instance 3' demanded a code distance of $25$. Thus, to avoid employing an additional $100$ qubits for `instance 1', it becomes crucial to determine the optimal code distance and the number of rounds needed. Since quantum computers are scarce and expensive, the users would be interested in optimizing the QEC parameters for obtaining desired computation results without paying for unnecessary resources. 

Identifying optimal parameters through simulations for surface codes is significantly time-consuming and complex. This is shown in Table \ref{tab:ibm_osaka_table} assuming a user conducts only $40$ trial-and-error for rounds.
For a single set of physical errors starting at the smallest distance, $d$ requires around $1.5$ hours for just one round. Considering the need to run about $40$ rounds for each distance to ensure accuracy, and assuming a user needs to determine parameters for $3$ different error models, the total time escalates dramatically. Specifically, for one set of physical errors, the cumulative simulation time reaches $160$ hours ($\sim6.6$ days).
The task of optimizing QEC parameters is especially challenging due to the need for rapid adjustments following regular calibrations. Our main challenge is quickly finding the optimal mix of code distance and rounds to meet a target logical error rate, considering the quantum system's current physical errors.


\textbf{Contribution:}
To the best of our knowledge, MITS is the first tool specifically tailored to account for the physical noise characteristics of quantum systems, to swiftly recommend a customized combination of distance and rounds for rotated surface codes. The aim is to achieve a target logical error rate while maintaining an optimal balance between qubit, gate, and time usage. The contributions to building MITS are as follows:
\circled{1}
\textit{Utilization of Quantum Simulation:} We employ STIM \cite{gidney2021stim}, a leading simulator for quantum stabilizer circuits. \textit{As indicated by a recent study} \cite{krinner2022realizing}\textit{, STIM effectively mirrors the performance of QECCs on real quantum machines.} Given the qubit limitations in quantum computers, making execution of full QECCs infeasible, STIM proves invaluable for our purposes.
\circled{2}
\textit{Development of MITS from STIM's Framework:} While STIM operates conventionally — using parameters like distance, rounds, and physical error to yield logical error rates — MITS adopts an inverse approach. We have tailored a dataset from STIM, aiming to infer the optimal distance and round configurations. This reverse engineering substantially expedites the process, cutting down simulation time.
\circled{3}
\textit{Model Selection using Heuristics and Machine Learning:} Recognizing that one-size-fits-all models are rarely effective, we explored multiple heuristics and machine learning models. This iterative approach allowed us to evaluate and select the best-fit model for our predictive needs, thereby ensuring that MITS's recommendations are reliable. 
\circled{4}
\textit{Flexibility of the Framework:} MITS is designed to be universally applicable, not exclusively tailored to STIM. It can be trained using any QECC simulator-generated or real hardware dataset. We assume readers are familiar with basic QEC concepts, including classical error correction, repetition codes, stabilizer \& encoding circuits, and surface codes. For such background information, one can refer to recent literature on the topic such as \cite{fowler2012surface}.

The tool, along with its code base, is available in a public GitHub repository. 
\footnote{GitHub Repository Link: \href{https://github.com/Avimita-amc8313/MITS-A-Quantum-Sorcerer-s-Stone-For-Designing-Surface-Codes}{MITS: A Quantum Sorcerer’s Stone For Designing Surface Codes}}

\textbf{Paper Structure:}
Section \ref{sec:background} overviews surface codes. Section \ref{sec:mits} discusses constructing MITS by reverse engineering STIM, creating datasets, and exploring various heuristics and machine learning models. Section \ref{sec:evaluation} compares and evaluates the selected model, demonstrating its effectiveness. The paper concludes in Section \ref{sec:conclusion}.
\section{Background on Surface Codes} \label{sec:background}


\subsection{Dynamics of Rotated Surface Codes}

\begin{figure}
    \centering
    \includegraphics[width=0.9\linewidth]{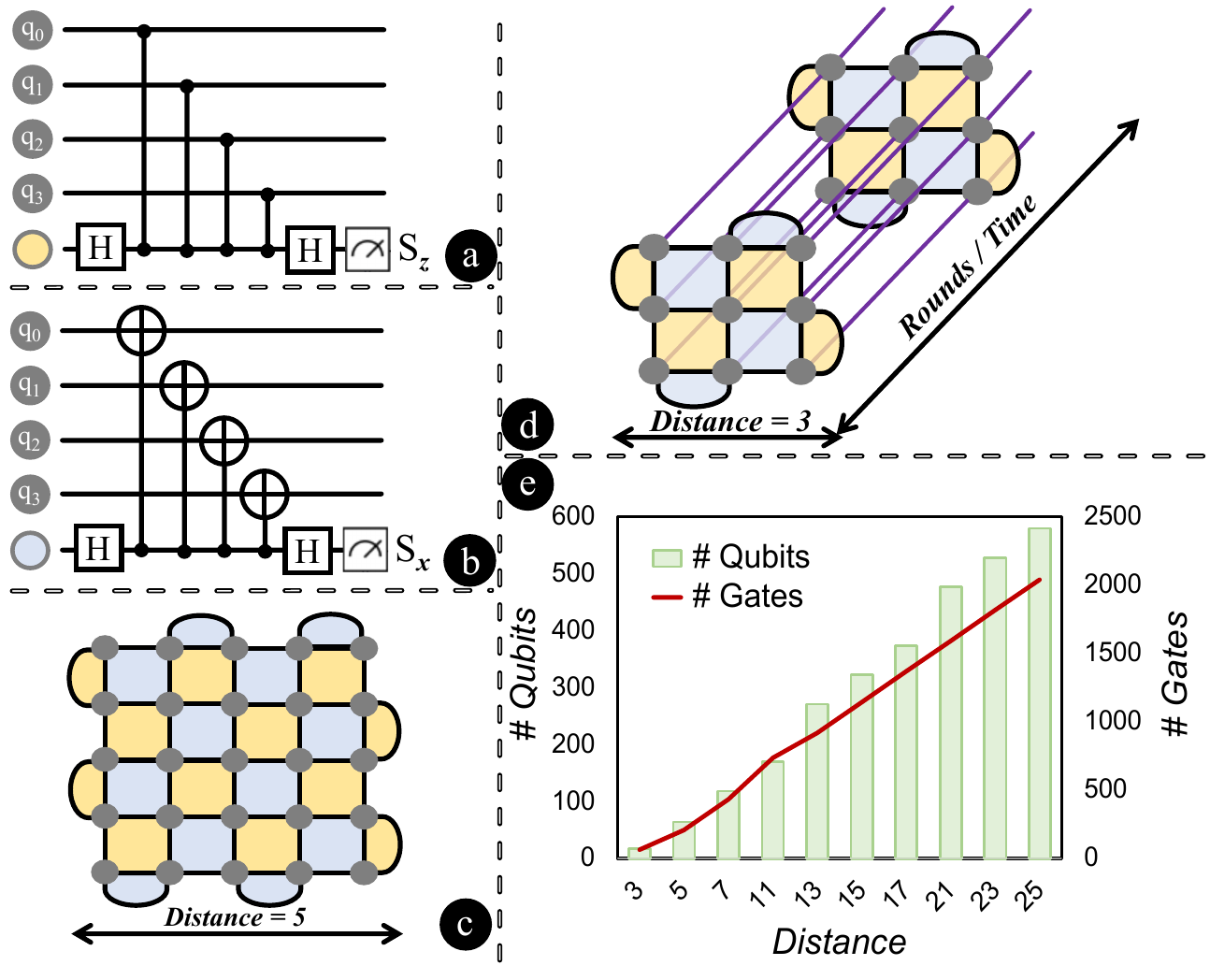}
    \caption{\textbf{Rotated surface codes: interactions, structures, and overheads.} 
    \raisebox{.5pt}{\textcircled{\raisebox{-.4pt} {\textbf{a}}}}
    Z-stabilizers interact with four data qubits. 
    \raisebox{.5pt}{\textcircled{\raisebox{-.9pt} {\textbf{b}}}}
    X-stabilizers interact with the same qubits. 
    \raisebox{.5pt}{\textcircled{\raisebox{-.2pt} {\textbf{c}}}}
    A distance $5$ rotated surface code forms a $5 \times 5$ lattice, with grey-blobs as data qubits and yellow/blue surfaces representing Z/X-stabilizers. 
    \raisebox{.5pt}{\textcircled{\raisebox{-.9pt} {\textbf{d}}}}
    A distance $3$ code across $2$ rounds highlights the $3 \times 3$ lattice's repetition over time. 
    \raisebox{.5pt}{\textcircled{\raisebox{-.4pt} {\textbf{e}}}}
    Increased distances raise qubit counts, and more rounds lead to greater gate numbers. With rounds being threefold the distance, the overhead of high distance and rounds is evident.
    }
    \label{fig:surface_background}
    \vspace{-20pt}
\end{figure}

At the heart of most QECCs lie two key stabilizers: the X-stabilizer and the Z-stabilizer. The role of X-stabilizers is to detect Z-flip or phase-flip errors, while Z-stabilizers identify X-flip or bit-flip errors. As illustrated in Fig. \ref{fig:surface_background} \raisebox{.5pt}{\textcircled{\raisebox{-.4pt} {\textbf{a}}}}, Z-stabilizers, when applied to four data qubits $(q_0, q_1, q_2, q_3)$, produce the syndrome $Z(q_0) \otimes Z(q_1) \otimes Z(q_2) \otimes Z(q_3)$. This syndrome is then projected onto an ancilla qubit, depicted as a yellow-blob. The resulting syndrome measurement, $S_z$, provides an indication of $\pm 1$ based on the presence of a bit-flip error in the data qubits. In a parallel fashion, X-stabilizers, as shown in Fig. \ref{fig:surface_background} \raisebox{.5pt}{\textcircled{\raisebox{-.9pt} {\textbf{b}}}}, interacting with the same set of data qubits, project the syndrome $X(q_0) \otimes X(q_1) \otimes X(q_2) \otimes X(q_3)$ onto the ancilla qubits, marked as a blue-blob. Here, the syndrome measurement, $S_x$, reveals whether there is a phase-flip error, with outcomes also manifesting as $\pm 1$.

The surface code operates on an $n \times n$ lattice foundation for an $n$-distance code, with the vertices in the lattice representing a data qubit. The lattice's design ensures that every qubit is under the surveillance of both X and Z stabilizers, with the mandate to detect phase and bit-flip errors. The presence of an error can alter the outcome of the associated stabilizer checks, indicating its occurrence.  As depicted in Fig. \ref{fig:surface_background} \raisebox{.5pt}{\textcircled{\raisebox{-.4pt} {\textbf{c}}}}, a distance $5$ rotated surface code is illustrated as a $5 \times 5$ lattice. Within this configuration, the grey-blobs denote data qubits. Each yellow surface signifies a Z-stabilizer, while the blue surfaces correspond to X-stabilizers. Utilizing a decoding algorithm, the location and nature of these errors can be pinpointed \cite{kolmogorov2009blossom, higgott2022pymatching}. Once identified, quantum gates intervene to rectify these errors, restoring the affected qubits to their rightful states. The strength of the surface code is its capability to amend multiple errors and its robustness amidst noise, as long as the noise level remains below a certain threshold.

The physical error rate reflects errors during individual qubit operations in a quantum computer, while the logical error rate gauges residual errors after applying QECC. The aim of QECCs is to minimize the logical error rate despite high physical errors. The `threshold' error rate represents the maximum physical error rate a QECC can handle while still reducing the logical error. Beyond this, the QECC becomes ineffective. Modern quantum processors manifest error rates close to $10^{-3}$ \cite{huang2019fidelity, foxen2020demonstrating}. Intriguingly, surface codes can tolerate an error threshold of approximately $10^{-2}$ \cite{krinner2022realizing, chatterjee2023q}, a value that is an order of magnitude higher than what current quantum processors can achieve without error correction.

\subsection{Interplay Between Code Distance and Rounds}

The distance of a surface code refers to the dimensions of the underlying lattice, with a distance $d$ corresponding to a $d \times d$ lattice. The round pertains to the number of times this lattice pattern is employed in the error correction process. With each round, the code attempts to identify and rectify errors, enhancing the probability of successful error correction, especially in the presence of persistent or recurrent errors. Essentially, for a distance $d$ surface code, the $d \times d$ lattice configuration is iterated for $r$ rounds. In Fig. \ref{fig:surface_background} \raisebox{.5pt}{\textcircled{\raisebox{-.9pt} {\textbf{d}}}}, a distance $3 \times 3$ rotated surface code is visualized across $2$ rounds, demonstrating how the $3 \times 3$ lattice pattern is repeated over time. The iteration over multiple rounds signifies the prolonged computational duration. It is evident that as the distance escalates, the total data qubits in the system also grow. Likewise, the gate count grows with an augmentation in the number of rounds. This relationship between distance and the increasing tally of qubits and gates is illustrated in Fig. \ref{fig:surface_background} \raisebox{.5pt}{\textcircled{\raisebox{-.4pt} {\textbf{e}}}}. For each distance value considered, the number of rounds is set as three times the distance. The depiction underscores the overhead associated with employing surface codes of increased distance and rounds.

Balancing distance and rounds in surface codes is pivotal to their efficacy. While it might seem tempting, from a theoretical standpoint, to lean towards a minimal distance surface code and ramp up the rounds, this approach encounters practical limitations. First, when the distance of a surface code is extended, there is an inherent need to also boost the rounds. The reason is that increased rounds add more gates, inadvertently introducing more errors. The expanded distance serves as a buffer, offering redundant checks to mitigate and rectify these errors. Secondly, the higher number of rounds inadvertently increases computational time. As a result, the data qubits are more susceptible to heat relaxation \& decoherence, introducing additional errors. Hence, the crux lies in meticulously calibrating the equilibrium between distance and rounds, ensuring optimal performance of the code.

\begin{figure}
    \centering
    \includegraphics[width=0.9\linewidth]{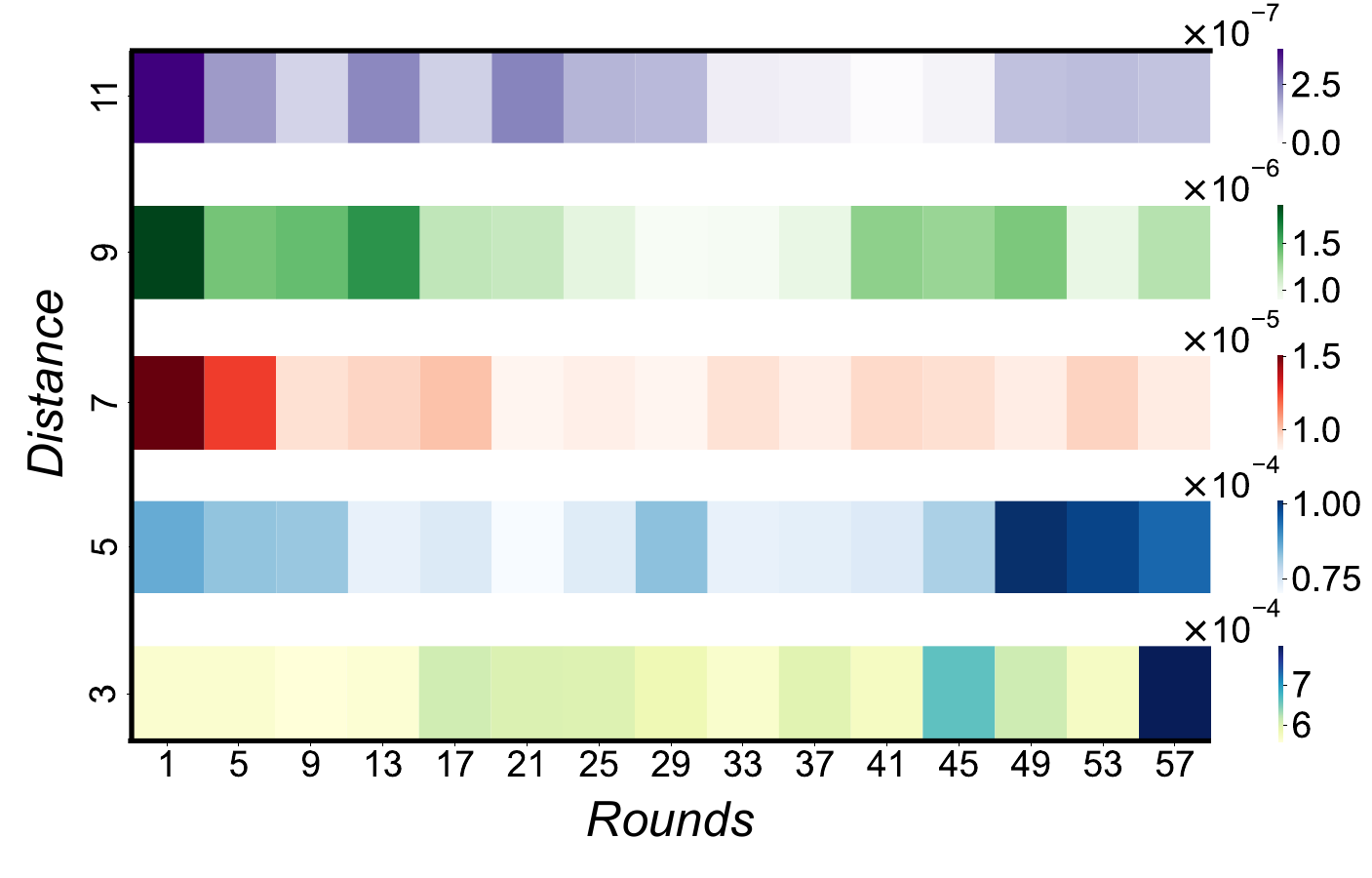}
    \caption{\textbf{Logical error rate heat map analysis.} With increasing distance, the logical error rate shows an anticipated reduction. Yet, there is a nuanced relationship with rounds: the error rate fluctuates initially, reaches a `sweet spot' where distance seems to counterbalance the errors from rounds sufficiently, and subsequently increases again, potentially from heat relaxation risks.}
    \label{fig:background_heat}
    \vspace{-15pt}
\end{figure}

In an experiment, we varied the distance and rounds of rotated surface codes and observed the resulting logical error rate (Fig. \ref{fig:background_heat}). As expected, an increased distance led to a reduced logical error rate. However, the interplay with rounds was intricate: the logical error rate fluctuated initially until it stabilized, suggesting the distance adequately counteracted the errors introduced by the rounds. This `sweet spot' state persisted for a few rounds before the logical error rate began rising again, likely due to the risk of heat relaxation \& decoherence. With a constant physical noise level, the challenge is pinpointing the perfect equilibrium between distance and the corresponding number of rounds to minimize qubit and gate usage.

\section{MITS: An Inverted Approach} \label{sec:mits}

\subsection{Pivotal Idea}
\begin{figure}
    \centering
    \includegraphics[width=0.9\linewidth]{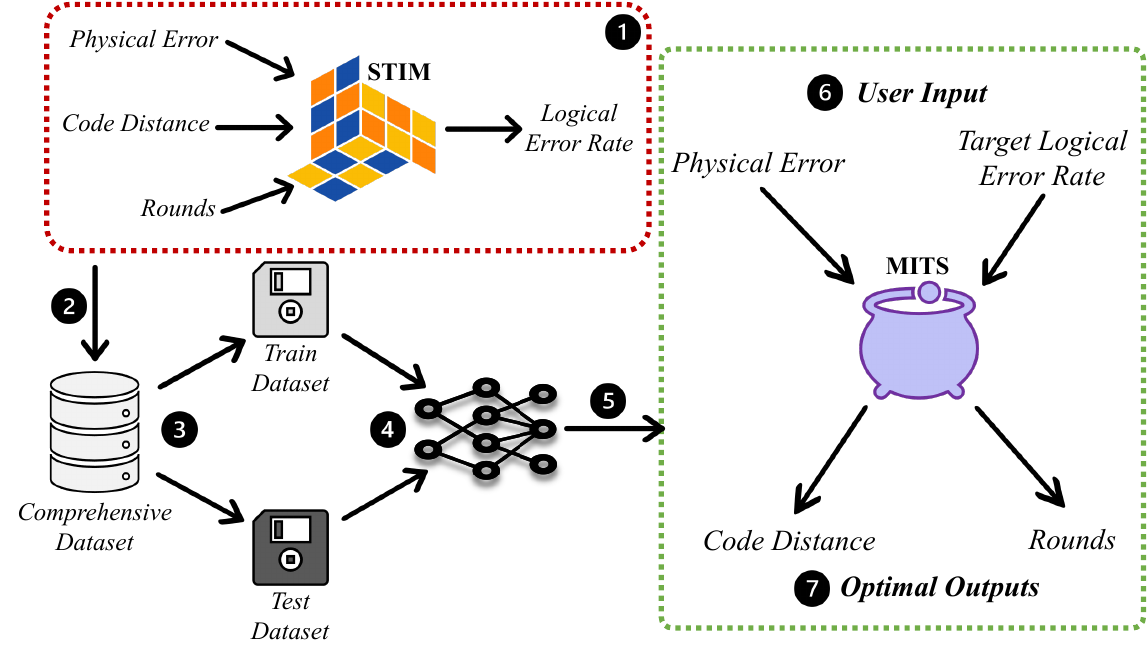}
    \caption{\textbf{The developmental and operational blueprint of MITS.} 
    \raisebox{.5pt}{\textcircled{\raisebox{-.9pt} {\textbf{1}}}}
    \textit{STIM Simulation:} Using STIM to estimate logical error rates from physical attributes, distance, and rounds.
    \raisebox{.5pt}{\textcircled{\raisebox{-.9pt} {\textbf{2}}}}
    \textit{Data Compilation:} Generated a dataset from $8.5k$ STIM experiments over weeks.
    \raisebox{.5pt}{\textcircled{\raisebox{-.9pt} {\textbf{3}}}}
    \textit{Dataset Partition:} Split into training and test subsets.
    \raisebox{.5pt}{\textcircled{\raisebox{-.9pt} {\textbf{4}}}}
    \textit{Model Exploration:} Evaluated various heuristic and ML models for optimal surface code parameters.
    \raisebox{.5pt}{\textcircled{\raisebox{-.9pt} {\textbf{5}}}}
    \textit{Optimal Model Selection:} Chose a two-step model using xgboost and random forest.
    \raisebox{.5pt}{\textcircled{\raisebox{-.9pt} {\textbf{6}}}}
    \textit{User Input:} Users specify physical error attributes and desired logical error rate.
    \raisebox{.5pt}{\textcircled{\raisebox{-.9pt} {\textbf{7}}}}
    \textit{Output:} MITS recommends optimal distance and rounds.
    }
    \label{fig:mits}
    \vspace{-15pt}
\end{figure}

If a user knows the noise level of their quantum machine (typically available for hardware) and has a target error rate in mind, they can input this information into MITS. In return, MITS will recommend the optimal `distance' and `rounds' for a rotated surface code. It will save hours the users would otherwise spend on simulations to refine distance and round using trial-and-error. We prepared a detailed dataset using STIM simulations and then employed various machine learning models and heuristics to predict the distance and rounds. This section first explains the dataset creation followed by testing and selection of the best model. Fig. \ref{fig:mits} illustrates the systematic progression of MITS, detailing both its development stages and operational workflow.

\subsection{Dataset Compilation from STIM}

\begin{figure*}[t]
    \centering
    \includegraphics[width=0.9\linewidth]{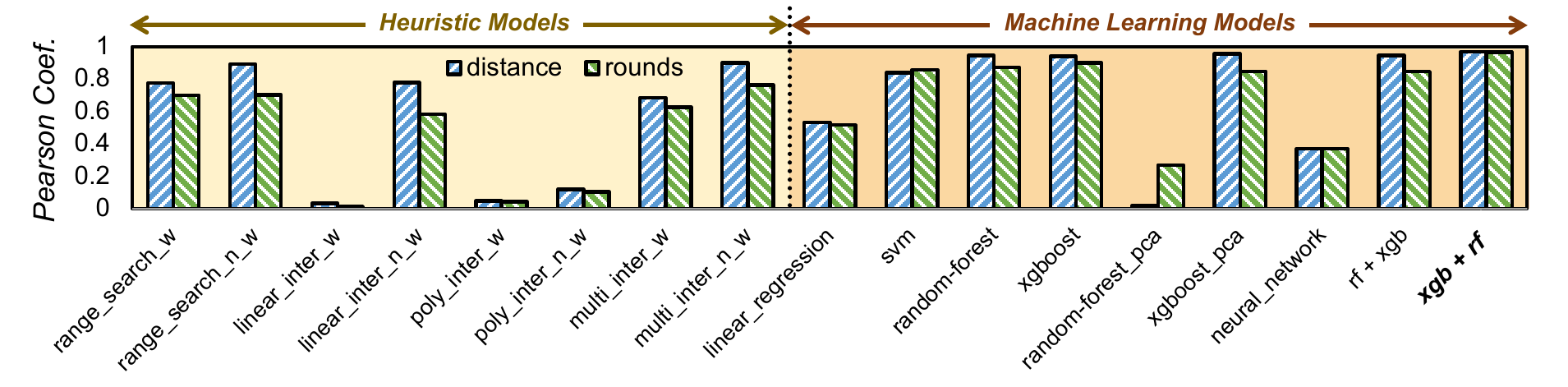}
    \caption{\textbf{Pearson Correlation of Heuristic vs. ML Models.} The bar graph shows heuristic models outperformed by XGBoost and Random Forest in machine learning. Optimal are XGBoost for distance and Random Forest for rounds, both in bold.}
    \label{fig:ablation}
    \vspace{-10pt}
\end{figure*}

Utilizing STIM \cite{gidney2021stim}, a stabilizer simulator, we estimated logical error rates based on physical error attributes, distance, and rounds of rotated surface codes. In total, we conducted $8640$ experiments, culminating in a database built over weeks. Our exploration focused on four types of physical error rates: depolarizing, gate, reset, and readout errors. \textit{We determined the error range for each type by examining the minimum and maximum errors across all available IBM quantum computers. This range is representative of most quantum computers and is subject to minimal variation due to calibration.} The experiments systematically spanned distances of $3$ to $19$ and rounds between $1$ to $60$. Notably, when the logical error rate reached or fell below $10^{-9}$ in our tests, STIM deems it fault-tolerant. Consequently, any subsequent trials with the current set of error attributes were terminated, preventing redundant additions to the dataset due to the known inverse relationship between error rates and code distance. Drawing from actual quantum computer error levels and carefully curated to exclude redundant values, this dataset stands as the most comprehensive resource tailored for this specific purpose. Across all experiments, we employed the Minimum Weight Perfect Matching (MWPM) decoder \cite{higgott2022pymatching}. For computational efficiency, the simulations utilized 4 parallel workers, enabling faster execution across multiple CPU cores. The maximum number of trials, or `shots', was set to 1,000,000, with each shot representing a single instance of the quantum error correction process. Post-data collection, we segregated the data, randomly reserving $20\%$ as a test set. Both datasets were pivotal for developing the predictive models that underpin MITS.

\subsection{Exploration for Predictive Models }

We started with heuristics, as they provide intuitive approaches, before delving into the complexities of machine learning to find the ideal model. 
We follow a two-step predictive method across all of our models: firstly predicting the code distance using the physical error rate and target logical error rate, and secondly predicting the rounds based on the deduced distance and target logical error rate. This approach is adopted for several reasons. The distance directly influences the logical error rate and serves as a foundational parameter of a surface code. By predicting the distance based on physical error rates and the desired logical error rate, we streamline the process, allowing for a more focused prediction. With the distance determined, the complexity of predicting rounds is reduced. If we aimed to predict both distance and rounds simultaneously, the inherent complexity would rise, potentially compromising the accuracy of our models. It is worth noting that when distance predictions are decimal, we round them up for better round predictions in the subsequent model. 
The Pearson correlation coefficient \cite{cohen2009pearson} measures the correlation between actual and predicted variables; a value near $0$ indicates little to no correlation, while a value near $1$ suggests a strong positive correlation. Fig. \ref{fig:ablation} showcases a comparison of all models evaluated during our exploration, using the Pearson coefficient as the metric.

\textbf{Heuristic Methods:}
We examined heuristics from two perspectives: assigning weights to physical error attributes based on their system impact, with gate errors receiving the highest due to their extensive effect, followed by depolarizing, readout, and reset errors. Weighted models (`\_w') use these aggregated weights as a singular feature, whereas non-weighted models (`\_n\_w') treat each error attribute independently without specific weights.
\circled{1}
\textit{Search by Range:} This model uses a distance-based heuristic to find the nearest points in a training dataset. It calculates the Euclidean distance between prediction data and training entries and then retrieves the associated code distance and rounds.
\circled{2}
\textit{Linear Interpolation:} This model performs linear interpolation on a training dataset to estimate distance and round values for a prediction dataset. Linear interpolation is a method of estimating values between two known data points using the formula: \( y = y_1 + (x - x_1) \times \frac{y_2 - y_1}{x_2 - x_1} \). The function sorts the training data based on proximity to each reference value from the prediction set and then uses the two nearest points for interpolation.
\circled{3}
\textit{Polynomial Interpolation:} Using a training dataset, this model fits a $2^{nd}$ degree polynomial to the three closest points for each prediction value. If the polynomial fit encounters an error, the function gracefully falls back to linear interpolation using the two nearest points.
\circled{4}
\textit{Multivariate Interpolation:} This function employs multivariate interpolation to estimate distance and round values for a prediction dataset based on a training dataset. For each entry in the prediction set, the function uses the `$griddata$' method to interpolate values based on the given input attributes. The training data serves as the input points and values for this interpolation. 

\textbf{Machine Leaning Models:}
Our objective was to ascertain whether the machine learning algorithms could offer improved predictive accuracy over the heuristic approaches.
\circled{1}
\textit{Linear Regression:} Using two sequential models, the first predicts a feature distance with hyperparameter tuning via grid search. The predictions then serve as input features for the second model, which predicts the optimal round.
\circled{2}
\textit{Support Vector Regression:} In a two-stage process, the first SVR model predicts a distance using hyperparameter tuning. Its predictions are then used as inputs for the second model, which predicts another target feature. Both stages use five-fold cross-validation for robustness.
\circled{3}
\textit{Random Forest Regression:} The first model predicts a distance based on several features and undergoes hyperparameter tuning. Its rounded predictions, combined with another feature, are inputs for the second model predicting rounds. A variant incorporating principal component analysis was also explored.
\circled{4}
\textit{XGBoost Regression:} In this two-stage process, the first model predicts a distance and undergoes hyperparameter tuning. Its rounded predictions are then used by the second model to predict rounds, with both stages using five-fold cross-validation.
\circled{5}
\textit{Neural Network:} A two-stage deep learning approach is employed. The first neural network predicts distance, and its predictions serve as input for the second model predicting rounds. Both models have three-layer architectures and use the Mean Squared Error loss with the Adam optimizer.
\section{Comparison and Evaluation} \label{sec:evaluation}

\begin{table}[]
\centering
\caption{Pearson Correlation Coefficients of Raw and Rounded Distance and Rounds of MITS}
\begin{tabular}{c|cc}
\hline \hline
\multirow{2}{*}{\textbf{Type}} & \multicolumn{2}{c}{\textbf{Pearson Corr. Coeff.}} \\ \cline{2-3} 
                               & \multicolumn{1}{c|}{\textbf{distance}}   & \textbf{rounds}  \\ \hline
\textbf{Raw Prediction}        & \multicolumn{1}{c|}{$0.982404$}            & $0.964948$         \\
\textbf{Rounded Prediction}    & \multicolumn{1}{c|}{$0.968062$}            & $0.964587$         \\ \hline \hline
\end{tabular}
\label{tab:pearson}
\vspace{-20pt}
\end{table}

After assessing various heuristics and machine learning models, multivariate interpolation stood out as the top heuristic. Both XGBoost and Random Forest showed similar high performance, as depicted in Fig. \ref{fig:ablation}. We chose one model for distance prediction and another for round prediction, capitalizing on their strengths. While other models could be viable for prediction, the exceptional performance of our chosen models negated the need to explore additional options. This section delves into the details of XGBoost and Random Forest's roles in distance and round predictions.

Predictive models, in their essence, do not always yield whole number outcomes. For distance, we round up the raw prediction to the subsequent odd number. While theoretically, surface codes can accommodate even distances, in practical scenarios, odd distances have proven to be more resilient and efficient. This is primarily because odd distances offer a better balance of error correction capability, ensuring the system's robustness. For rounds, we adjust the raw predictions to the nearest whole number. Regardless of the raw value for distance or rounds, we consistently elevate it to the next odd or whole number. This strategy is imperative to consistently achieve the desired logical error rate. Reducing the distance could jeopardize meeting this target rate. Both raw and adjusted predictions are vital in our analysis. In our approach to finalize the MITS, we employed a two-tiered modeling strategy. Initially, we utilized the XGBoost's $XGBRegressor$ for predicting distance. Key parameters for this model included a \textit{learning\_rate} of $0.1$, \textit{max\_depth} of $6$, \textit{min\_child\_weight} of $5$, \textit{gamma} of $0.5$, and \textit{n\_estimators} set to $200$. Following this, the predicted distance values were fed into a Random Forest Regressor to predict rounds. The Random Forest model was configured with a \textit{max\_depth} of $20$, \textit{min\_samples\_split} of $10$, and \textit{n\_estimators} set to $10$. The training times for the models were approximately $30$ minutes and $7$ minutes, respectively. The shorter duration for the latter is due to its use of a smaller dataset containing only distance and target logical error rate attributes, compared to the former which utilized the full dataset. Table \ref{tab:pearson} displays the Pearson correlation coefficients of these models with respect to raw and rounded values. Hyperparameter tuning with $GridSearchCV$ was utilized to prevent overfitting. \textit{Validation on the training dataset yielded correlation coefficients of $0.986$ and $0.967$ for models predicting $d$ and $r$, closely matching Pearson coefficients from the test dataset predictions. This similarity confirms the absence of overfitting.}

\begin{figure}
    \centering
    \includegraphics[width=0.9\linewidth]{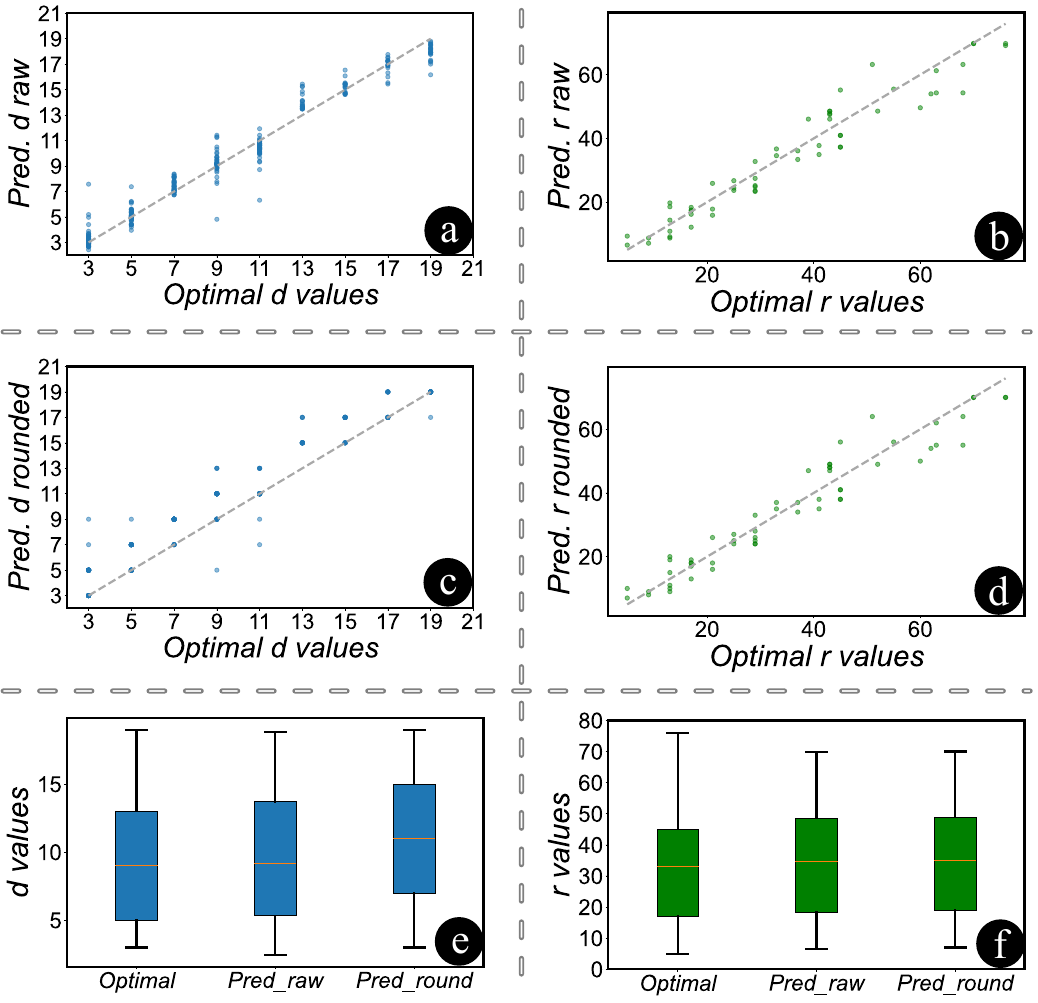}
    \caption{\textbf{Comparative Analysis of Predicted vs. Optimal Distance and Rounds.} \raisebox{.5pt}{\textcircled{\raisebox{-.4pt} {\textbf{a}}}} and \raisebox{.5pt}{\textcircled{\raisebox{-.4pt} {\textbf{c}}}}: Scatter plots comparing predicted and optimal distances, showing a trend towards rounding to the nearest odd number starting at three.
    \raisebox{.5pt}{\textcircled{\raisebox{-.9pt} {\textbf{b}}}} and \raisebox{.5pt}{\textcircled{\raisebox{-.9pt} {\textbf{d}}}}: Scatter plots of rounds, with raw values generally rounding up to the nearest integer, showing minimal variance. Dotted diagonals indicate perfect predictions. 
    \raisebox{.5pt}{\textcircled{\raisebox{-.4pt} {\textbf{e}}}} and \raisebox{.5pt}{\textcircled{\raisebox{-.9pt} {\textbf{f}}}}: Box plots contrasting original, predicted, and rounded values. Distances show deviations upon rounding, while rounds remain consistent across all measures.}
    \label{fig:best_results}
    \vspace{-10pt}
\end{figure}

The scatter plots in Fig. \ref{fig:best_results} \raisebox{.5pt}{\textcircled{\raisebox{-.4pt} {\textbf{a}}}} and \raisebox{.5pt}{\textcircled{\raisebox{-.4pt} {\textbf{c}}}} compare the predicted distance values, both in their raw form and when rounded, to the optimal distance values. It is evident from the plots that all distance values are rounded up to the nearest odd number, with a minimum value set at three. In Fig. \ref{fig:best_results} \raisebox{.5pt}{\textcircled{\raisebox{-.9pt} {\textbf{b}}}} and \raisebox{.5pt}{\textcircled{\raisebox{-.9pt} {\textbf{d}}}}, scatter plots are depicted comparing rounds in their raw form and when rounded. Given that raw rounds are simply rounded up to the nearest integer, there is minimal discernible difference between the two plots. The intricacies and variations among the original, predicted, and rounded values for both distance and rounds can be more comprehensively understood through the box plots. Fig. \ref{fig:best_results} \raisebox{.5pt}{\textcircled{\raisebox{-.4pt} {\textbf{e}}}} and \raisebox{.5pt}{\textcircled{\raisebox{-.9pt} {\textbf{f}}}} provide a clear visual representation of these values. Upon examination of the data, it is evident that the original and predicted values for distance are closely aligned. However, a noticeable divergence emerges when the predicted values are rounded up to the nearest odd integer. Conversely, for the rounds, there is no discernible difference between the original, predicted, and rounded values. The Pearson correlation coefficients in Table \ref{tab:pearson} demonstrate that, despite rounding both the distances and rounds, the predicted results remain closely aligned with the actual values.

\begin{figure}
    \centering
    \includegraphics[width=0.9\linewidth]{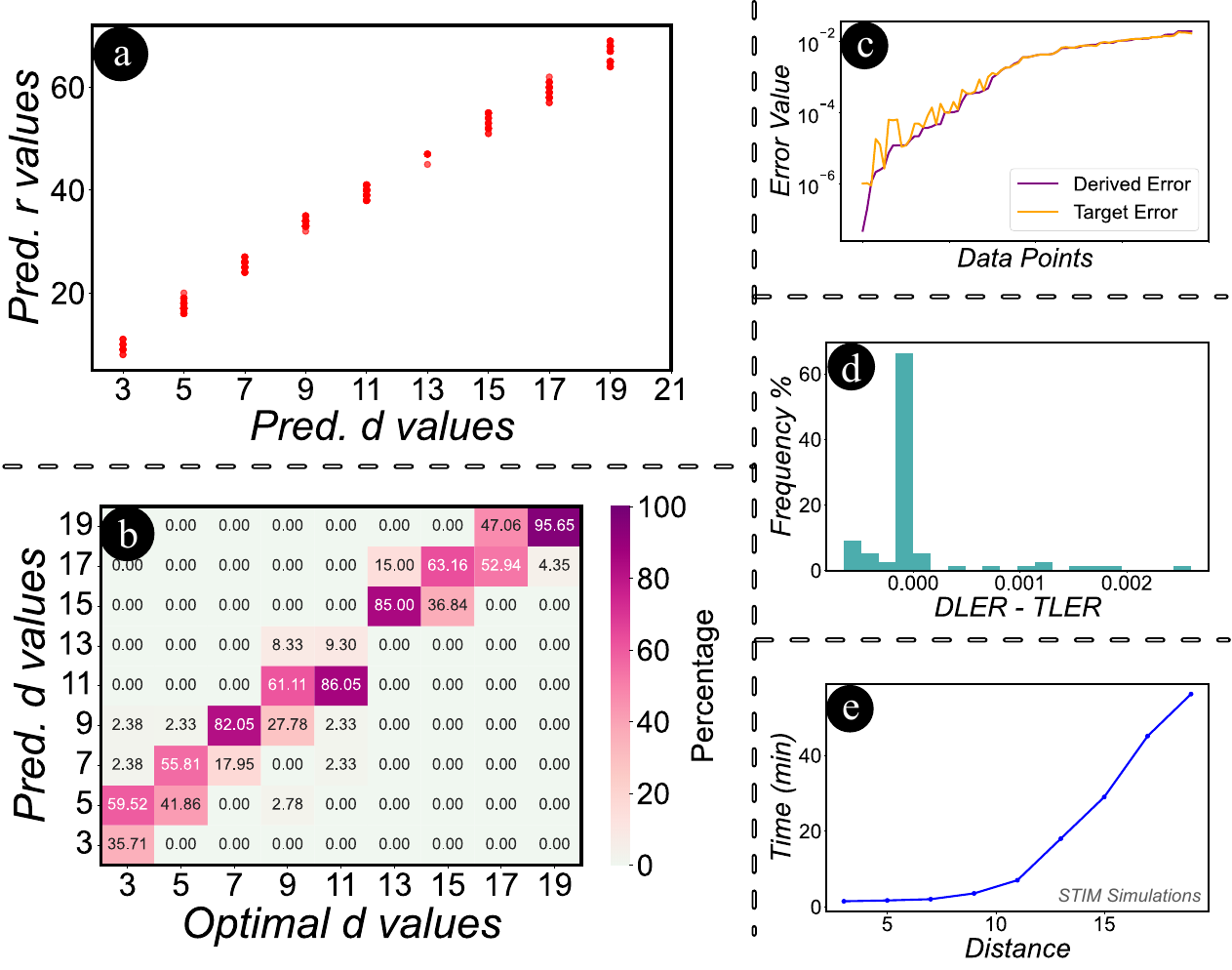}
    \caption{\textbf{Analysis of Predictions, Error Rates, \& Simulation Efficiency.} 
    \raisebox{.5pt}{\textcircled{\raisebox{-.4pt} {\textbf{a}}}}: Scatter plot showcasing the correlation between predicted distance and rounds.
    \raisebox{.5pt}{\textcircled{\raisebox{-.9pt} {\textbf{b}}}}: Heatmap of predicted vs. optimal d values.
    \raisebox{.5pt}{\textcircled{\raisebox{-.4pt} {\textbf{c}}}}: Relationship between derived and target error.
    \raisebox{.5pt}{\textcircled{\raisebox{-.9pt} {\textbf{d}}}}: Histogram highlighting MITS's accuracy with a centered distribution at $0$ for the difference between DLER and TLER. MITS trends towards reducing logical error rate with minimal negative deviations.
    \raisebox{.5pt}{\textcircled{\raisebox{-.4pt} {\textbf{e}}}}: STIM simulation time growth with increasing d and r, underscoring MITS’s efficiency in reducing time to around $11 \pm 3$ milliseconds.
    }
    \label{fig:error_of_error}
    \vspace{-10pt}
\end{figure}

Next, we evaluate the logical error rates. To begin with a recap, it is essential to note that for surface codes to yield optimal results, there is a necessity to increase the rounds in tandem with the distance. Fig. \ref{fig:error_of_error} \raisebox{.5pt}{\textcircled{\raisebox{-.4pt} {\textbf{a}}}} illustrates a scatter plot comparing the predicted distance and rounds. The evident trend of increasing rounds with the rise in distance underscores our assertion.
In Fig. \ref{fig:error_of_error} \raisebox{.5pt}{\textcircled{\raisebox{-.9pt} {\textbf{b}}}}, we present a heatmap illustrating the comparison between predicted d values and their optimal counterparts. This visualization aids in determining the frequency with which certain values are predicted as d. Particularly, we focus on instances where predictions fall below the optimal values, as these cases signify potential failures to achieve the target logical error rate.
We performed another experiment to validate the logical error rate obtained from STIM by using the $d$ and $r$ values predicted by MITS and comparing them against the target logical error rate. 
In Fig. \ref{fig:error_of_error} \raisebox{.5pt}{\textcircled{\raisebox{-.4pt} {\textbf{c}}}}, the relationship between the derived error and the target error is illustrated. Notably, the derived error rate is predominantly lower than the target rate, with a few exceptions exhibiting minimal discrepancies.
We quantify the discrepancy between the Derived Logical Error Rate (DLER) and the Target Logical Error Rate (TLER) by subtracting them. Fig. \ref{fig:error_of_error} \raisebox{.5pt}{\textcircled{\raisebox{-.9pt} {\textbf{d}}}} presents a histogram illustrating the frequency distribution of this difference. From the histogram, it is evident that the predominant distribution is centered at $0$, affirming that MITS consistently achieves the target logical error rate. A significant portion of the distribution also leans towards the positive side, indicating that MITS often surpasses the target by reducing the logical error rate even further — a favorable outcome. There is a minor distribution on the negative spectrum, the magnitude of these differences is so minuscule that they can be considered negligible. 
Figure \ref{fig:error_of_error} \raisebox{.5pt}{\textcircled{\raisebox{-.4pt} {\textbf{e}}}} depicts the exponential increase in STIM's simulation time for surface codes as distance $d$ and rounds $r$ grow. Adjusting for rounds and variable physical error rates requires more time.
Given the daily error calibrations of quantum computers, users may frequently need to adjust parameters to maintain efficiency. This task can prove particularly challenging when dealing with multiple variations in physical errors and target outcomes, significantly increasing the time required to identify optimal parameters.
MITS significantly reduces trial-and-error based simulation time from hours to $11 \pm3$ milliseconds on a pre-trained model, swiftly predicting the required distance and rounds to achieve a specified target logical error rate.

\section{Conclusion} \label{sec:conclusion}

This paper presented MITS, a new methodology to optimize surface code implementations by predicting ideal distance and rounds given target logical error rates and known physical noise levels of the hardware. By devising an inverse modeling approach and training on a comprehensive simulation dataset of over $8500$ experiments, MITS can rapidly recommend surface code parameters that balance qubit usage with error rate goals. Our comparative assessment validated the efficacy of the XGBoost and Random Forest models underpinning MITS's predictions, which achieved Pearson correlation coefficients of $0.98$ and $0.96$ for distance and rounds respectively. We further confirmed that the predicted combination of distance and rounds from MITS consistently attained the target logical error rates, with deviations centered around $0$. 
MITS can cut hours from surface code calibration, assisting in the realization of practical error-corrected quantum processors. 
\footnote{GitHub Repository Link: \href{https://github.com/Avimita-amc8313/MITS-A-Quantum-Sorcerer-s-Stone-For-Designing-Surface-Codes}{MITS}}

\bibliographystyle{unsrt}
\bibliography{refs}

\end{document}